\documentstyle[fleqn,12pt]{ioplppt}
\jl{1}
\eqnobysec
\begin{document}
\title{Coherent states for a quantum
particle on a circle}
\author{K Kowalski\dag, J Rembieli\'nski\dag\ and L C Papaloucas\ddag}
\address{\dag\ Department of Theoretical Physics, University
of \L\'od\'z, ul.\ Pomorska 149/153,\\ 90-236 \L\'od\'z,
Poland}
\address{\ddag\ Department of Mathematics, University of
Athens, Panepistomiopolis, 157 84\\ Athens, Greece}
\begin{abstract}
The coherent states for the quantum particle on
the circle are introduced.  The Bargmann representation
within the actual treatment provides the representation of
the algebra $[\hat J,U]=U$, where $U$ is unitary, which is a
direct consequence of the Heisenberg algebra $[\hat \varphi,
\hat J]=i$, but it is more adequate for the study of the
circlular motion.
\end{abstract}
\vspace{2.2cm}
\noindent Key words:\hspace{1cm}quantum mechanics, coherent
states, quantum groups

\vspace{2.2cm}
\pacs{02.10, 02.20, 02.30, 02.40, 03.65}
\maketitle
\newpage
\section{Introduction}
In spite of their importance the theory of quantization of
systems involving periodic motion seems to be far from
complete.  For instance, the difficulties has been recently
reported with quantization of the plane pendulum [1].  Namely,
different results were obtained depending whether the
constraints were imposed before or after the quantization.
Another example is yet unsolved problem of quantization of
the double pendulum [2].  We also recall the troublesome
questions [3] arising in the analysis of the Heisenberg
uncertainty relations arising in the case when one of
observables, as for instance the angle operator, has the
compact spectrum.

The experience with the harmonic oscillator suggests that the
concept of coherent states would be an important tool for
the better understanding of the periodic motion of a quantum
particle.  Furthermore, it is natural to restrict to the
simplest case of the quantum particle moving in a circle.

The purpose of this work is to introduce the coherent states
for the quantum particle on a circle.  As a matter of fact
there exists a method for generating coherent states for the
quantum particle moving in a circle from the standard
coherent states [4].  The obtained coherent states are
different from those defined in this paper.  Nevertheless,
it seems to us that the approach presented herein is more
adequate one.

In section 2 we introduce the algebra $[\hat J,U]=U$, where
$\hat J$ is the angular momentum operator and $U$ is
unitary.  This algebra is implied by the Heisenberg algebra
$[\hat\varphi,\hat J]=i$, where $\hat\varphi$ is the angle operator,
but it seems to be more adequate in the case with the
circular motion of a quantum particle.  In section 3 we
define the abstract coherent states and discuss their basic
properties.  Section 4 is devoted to the Bargmann
representation within the actual treatment.  This
representation provides us with the concrete realization of
the algebra $[\hat J,U]=U$, in the space of functions
possessing Laurent series expansion.  In section 5 we
illustrate the actual treatment by an example of the free
motion of the quantum particle in a circle.
\section{Quantum mechanics on the circle}
The purpose of this section is to discuss the fundamental
commutation relations and abstract Hilbert space of states
for the quantum particle on the circle.  In order to clarify
the notation we now specialize to the case of the free
motion in a circle.  The classical Lagrangian is given by
\begin{equation}
L = \frac{1}{2}\dot \varphi^2,
\end{equation}
where we have assumed that the particle has the unit mass
and it moves in the unit circle.  The angular momentum
canonically conjugate to the angle $\varphi$ is
\begin{equation}
J = \frac{\partial L}{\partial \dot \varphi}=\dot \varphi,
\end{equation}
so that the Hamiltonian can be written as
\begin{equation}
H = \frac{1}{2}J^2.
\end{equation}
Clearly, we have the Poissson bracket such that
\begin{equation}
\{\varphi,J\} = 1,
\end{equation}
which leads accordingly to the rules of the canonical
quantization to the commutator
\begin{equation}
[\hat\varphi,\hat J] = i,
\end{equation}
where we set $\hbar=1$.  The commutator (2.5) needs very
subtle analysis [3,5,6] involving common domain of operators $\hat
\varphi$ and $\hat J$.  On the other hand, the experience
shows that the study of the circular motion of
the quantum particle based on the formula (2.5) is
technically complicated.  It seems to us that the reason of
difficulties with the application of the commutator (2.5) is
the inadequate choice of $\hat \varphi$ as the position
observable for the quantum particle on a circle.  Indeed,
this observable arising from utilization of the usual
position operator for the particle on the real line does not
take into account the topology of the circle.  In our
opinion the good candidate to represent the position of the
quantum particle on the (unit) circle is the unitary
operator $U$ such that
\begin{equation}
U = e^{i\hat\varphi}.
\end{equation}
In fact, the spectrum of $U$ should obviously coincide with the
unit circle.  It should also be noted that $U$ is determined
on the whole Hilbert space of states.  Now, taking into
account (2.5) and (2.6) we arrive at the following algebra:
\begin{equation}
[\hat J,U] = U,
\end{equation}
where $U$ is unitary.  It is easy to verify with the help of
(2.7) that the spectrum of $U$ really coincides with the
unit circle.

Consider the eigenvalue equation
\begin{equation}
\hat J|j\rangle = j|j\rangle.
\end{equation}
From (2.8) and (2.7) it follows that the operators $U$ and
$U^\dagger $ act on the vectors $|j\rangle$ as the rising
and lowering operators, respectively, i.e. we have
\numparts
\begin{eqnarray}
U|j\rangle &=& |j+1\rangle,\\
U^\dagger |j\rangle &=& |j-1\rangle.
\end{eqnarray}
\endnumparts
In view of (2.9) we can generate the whole basis
$\{|j\rangle\}$ of the Hilbert space of states from the
unique vector (vacuum vector) $|j_0\rangle$, where
$j_0\in [0,1]$.  Clearly, different $j_0$ lead to
nonequivalent irreducible representations of the commutation
relations (2.7).  We now demand the time-reversal invariance
of the algebra (2.7), that is
\begin{eqnarray}
T\hat JT^{-1} &=& -\hat J,\\
TUT^{-1} &=& U^{-1},
\end{eqnarray}
where $T$ is the anti-unitary operator of time inversion.
The relations (2.8)--(2.11) taken together yield
\begin{equation}
T|j\rangle = |-j\rangle.
\end{equation}
Hence, we find that $T$ is well-defined on the Hilbert space
of states generated by the vectors $|j\rangle$ if and only
if the spectrum of $\hat J$ is symmetric with respect to
zero.  By virtue of (2.9) this means that the only
possibility left is $j_0=0$ or $j_0=\frac{1}{2}$.
Evidently, $j_0=0$ ($j_0=\frac{1}{2}$) implies integer
(half-integer) eigenvalues $j$.  We have thus shown by
demanding the invariance of (2.7) under the time inversions
which seems to be reasonable assumption provided we do not
take into consideration dissipative systems, that $j$ can be
only integer or half-integer.  Notice, that the most natural
interpretation of the case with integer (half-integer) $j$
is that it refers to bosons (fermions).  The fact that
solutions to the Schr\"odinger equation for the quantum
pendulum involve spin-half eigenfunctions was recognized by
Pradhan \etal [7].  We also remark that the subalgebra (2.7) of
the full algebra (2.7), (2.10) and (2.11) was postulated for the study
of quantum mechanics on the circle in [8].  Nevertheless, the
physical interpretation of the infinite number of
nonequivalent irreducible representations of (2.7) arising when
one restricts to (2.7) was not provided therein.  On the other
hand, it seems that the quantization procedure based solely
on (2.7) which does not preserve such the fundamental symmetry
of the free motion of the classical particle in the circle
as the time-reversal one is not physically acceptable.  We finally write
down the orthogonality
and completeness conditions satisfied by the vectors
$|j\rangle$ such that
\begin{eqnarray}
\langle j|k\rangle &=& \delta_{jk},\\
\sum_{j=-\infty}^{\infty} |j\rangle\langle j|&=& I
\end{eqnarray}
(substitution $j\to j-\frac{1}{2}$ in the case with fermions
understood).
\section{Abstract coherent states}
\subsection{Definition of coherent states}
In this section we introduce the abstract coherent states
for the quantum mechanics on the circle.  We begin with the
definition of these states.  Since the generators $U$ and
$U^\dagger $ of the algebra (2.7) are non-Hermitian (they
are unitary), therefore we cannot adopt the usual technique
of Perelomov [9] for generating coherent states.  Instead, we
follow the Barut-Girardello approach [10] and we seek the
coherent states as the solution of the eigenvalue equation
\begin{equation}
X|\xi\rangle = \xi|\xi\rangle,
\end{equation}
with complex $\xi$.  What is $X$ ?  Motivated by the form of
the eigenvalue equation satisfied by the standard coherent
states $|z\rangle$, where $z$ is complex, such that
\begin{equation}
e^{ia}|z\rangle = e^{iz}|z\rangle,
\end{equation}
where $a\sim \hat q+i\hat p$ is the standard Bose
annihilation operator and $\hat q$ and $\hat p$ are the
position and momentum observables, respectively, we put
\begin{equation}
X := e^{i(\hat \phi + i\hat J)}.
\end{equation}
Hence, using (2.6) and the Baker-Hausdorff formula
we find
\begin{equation}
X = Ue^{-\hat J - \hbox{$\frac{1}{2}$}}.
\end{equation}
A straightforward calculation shows that the operators $X$
and $X^\dagger $ are the ladder operators.  Namely,
\numparts
\begin{eqnarray}
X|j\rangle &=& e^{-j-\hbox{$\frac{1}{2}$}}|j+1\rangle,\\
X^\dagger |j\rangle
&=& e^{-j+\hbox{$\frac{1}{2}$}}|j-1\rangle.
\end{eqnarray}
\endnumparts
It should also be noted that the operator $X$ transforms
under the time inversions as (see (3.4), (2.10) and
(2.11)):
\begin{equation}
TXT^{-1} = X^{-1}.
\end{equation}
Now, expanding the coherent state $|\xi\rangle$ in the basis
$|j\rangle$
\begin{equation}
|\xi\rangle = \sum_{j=-\infty}^{\infty}c_j(\xi)|j\rangle,
\end{equation}
where $c_j(\xi)=\langle j|\xi\rangle$, and taking into
account (3.1) and (3.5a) we obtain the following recursive
relation:
\begin{equation}
c_{j+1} = \xi^{-1}e^{-j-\hbox{$\frac{1}{2}$}}c_j,
\end{equation}
where $j$ is integer.  Solving this elementary recurrence we
get
\begin{equation}
c_j = c_0(\xi)\xi^{-j}e^{-\frac{j^2}{2}},
\end{equation}
where $c_0(\xi)$ is an arbitrary function of $\xi$.  For the
sake of simplicity we shall set in the sequel
$c_0(\xi)\equiv 1$.  Therefore,
\begin{equation}
\langle j|\xi\rangle = \xi^{-j}e^{-\frac{j^2}{2}},
\end{equation}
and the coherent state $|\xi\rangle$ is given by
\begin{equation}
|\xi\rangle =
\sum_{j=-\infty}^{\infty}\xi^{-j}e^{-\frac{j^2}{2}}|j\rangle.
\end{equation}
We note that in accordance with usual properties of coherent
states the complex numbers $\xi$ should parametrize the
cylinder which is the classical phase space for the particle
moving in a circle.  Thus we can set
\begin{equation}
\xi = e^{-l + i\varphi}.
\end{equation}
Indeed, consider the cylinder
\begin{equation}
x=\cos\varphi,\qquad y=\sin\varphi,\qquad z=l.
\end{equation}
In order to identify the points of the cylinder with the
(compactified) complex plane one should to deform it and
then project the points of the obtained surface on the
$(x,y)$ plane.  It seems that the simplest possibility left
is to utilize the following transformation:
\begin{equation}
x'=e^{-z}x,\qquad y'=e^{-z}y,\qquad z'=z,
\end{equation}
which leads to the representation (3.12).  The
correspondence between the parameters $l$, $\varphi$ and the
classical angular momentum and angle variables,
respectively, is discussed in the next subsection.  We
observe that the singularity in (3.11) corresponding to
$\xi=0$ arises by virtue of (3.12) only for asymptotical
values of $l$.  It should be noted that the operator
$X^{-1}$ acts on the coherent states as follows:
\begin{equation}
X^{-1}|\xi\rangle = \xi^{-1}|\xi\rangle.
\end{equation}
We also write down the relation describing transformation of
the coherent states under the time inversions such that
\begin{equation}
T|\xi\rangle = |\xi^{*-1}\rangle,
\end{equation}
where the asterisk designates the complex conjugation.

In view of (3.12) we can write the coherent state in the
form
\begin{equation}
|l,\varphi\rangle =
\sum_{j=-\infty}^{\infty}e^{lj-ij\varphi}e^{-\frac{j^2}{2}}
|j\rangle,
\end{equation}
where $|l,\varphi\rangle\equiv|\xi\rangle$ with
$\xi=e^{-l+i\varphi}$.  Evidently, we have
\begin{equation}
\langle j|l,\varphi\rangle =
e^{lj-ij\varphi}e^{-\frac{j^2}{2}}.
\end{equation}
We remark that the coherent states obey
\begin{equation}
e^{-\eta\hat J}|\xi\rangle = |e^\eta\xi\rangle.
\end{equation}
Therefore, we can represent the state $|\xi\rangle$ as
\begin{equation}
|\xi\rangle = e^{-(\ln\xi) \hat J}|1\rangle,
\end{equation}
where by virtue of (3.11)
\begin{equation}
|1\rangle =
\sum_{j=-\infty}^{\infty}e^{-\frac{j^2}{2}}|j\rangle.
\end{equation}
In the parametrization $l$, $\varphi$ the above formulas
take the form
\begin{equation}
e^{(h-i\psi)\hat J}|l,\varphi\rangle =
|l+h,\varphi+\psi\rangle.
\end{equation}
Hence,
\begin{equation}
|l,\varphi\rangle = e^{(l-i\varphi)\hat J}|0,0\rangle,
\end{equation}
where $|0,0\rangle\equiv |1\rangle$ (see (3.12) and (3.17)).
The relations (3.19) and (3.22) follow directly from (2.8)
and (3.11).  Observe that the vector $|1\rangle$ is
invariant under the action of the operator $T$ (see (3.16)).

The coherent states are not orthogonal.  In fact, using
(3.10) and (3.11) we find
\numparts
\begin{eqnarray}
\langle \xi|\eta\rangle &=&
\sum_{j=-\infty}^{\infty}(\xi^*\eta)^{-j}e^{-j^2} =
\theta_3(\hbox{$\frac{i}{2\pi}$}\ln\xi^*\eta|\hbox{$\frac{i}
{\pi}$}),\qquad
\hbox{(boson case)},\\
\langle \xi|\eta\rangle &=&
\sum_{j=-\infty}^{\infty}(\xi^*\eta)^{-j+\frac{1}{2}}
e^{-(j-\frac{1}{2})^2} =
\theta_2(\hbox{$\frac{i}{2\pi}$}\ln\xi^*\eta|\hbox{$\frac{i}
{\pi}$}),\qquad
\hbox{(fermion case)}.
\end{eqnarray}
\endnumparts
Accordingly, in the parametrization (3.12) and (3.17)
\numparts
\begin{eqnarray}
\langle l,\varphi|h,\psi\rangle &=&
\theta_3(\hbox{$\frac{1}{2\pi}$}(\varphi-\psi)-\hbox{$\frac{l+h}{2}\frac{i}{
\pi}$}|\hbox{$\frac{i}{\pi}$}),\qquad \hbox{(boson case)}\\
\langle l,\varphi|h,\psi\rangle &=&
\theta_2(\hbox{$\frac{1}{2\pi}$}(\varphi-\psi)-\hbox{$\frac{l+h}{2}\frac{i}{
\pi}$}|\hbox{$\frac{i}{\pi}$}).\qquad \hbox{(fermion case)}
\end{eqnarray}
\endnumparts
The functions $\theta_3$ and $\theta_2$ in the above
formulas are the Jacobi theta-functions defined by
\numparts
\begin{eqnarray}
\theta_3(v|\tau) &=&
\sum_{n=-\infty}^{\infty}q^{n^2}(e^{i\pi v})^{2n},\\
\theta_2(v|\tau) &=&
\sum_{n=-\infty}^{\infty}q^{(n-\frac{1}{2})^2}(e^{i\pi
v})^{2n-1},
\end{eqnarray}
\endnumparts
where $q=e^{i\pi\tau}$ and $\hbox{Im}\,\tau>0$.  We note that
\begin{equation}
\theta_3(-v)=\theta_3(v),\qquad \theta_2(-v)=\theta_2(v).
\end{equation}
The introduced coherent states are not
normalized.  Indeed, it follows immediately from (3.24) and
(3.25) that
\numparts
\begin{eqnarray}
\langle \xi|\xi\rangle &=&
\theta_3(\hbox{$\frac{i}{\pi}$}\ln|\xi||\hbox{$\frac{i}{\pi}$}),
\qquad \hbox{(boson case)},\\
\langle \xi|\xi\rangle &=&
\theta_2(\hbox{$\frac{i}{\pi}$}\ln|\xi||\hbox{$\frac{i}{\pi}$}),
\qquad \hbox{(fermion case)}
\end{eqnarray}
\endnumparts
and
\numparts
\begin{eqnarray}
\langle l,\varphi|l,\varphi\rangle &=&
\sum_{j=-\infty}^{\infty}e^{2lj}e^{-j^2}=\theta_3
(\hbox{$\frac{il}{\pi}|\hbox{$\frac{i}{\pi}$}$}),\qquad
\hbox{(boson case)}\\
\langle l,\varphi|l,\varphi\rangle &=&
\sum_{j=-\infty}^{\infty}e^{2l(j-\frac{1}{2})}e^{-(j-\frac{1}
{2})^2}=\theta_2
(\hbox{$\frac{il}{\pi}|\hbox{$\frac{i}{\pi}$}$}).\qquad
\hbox{(fermion case)}
\end{eqnarray}
\endnumparts
Let finally recall that nonorthogonality is one of the
characteristic features of coherent states.
\subsection{Coherent states and the classical phase space}
As we have promised in the previous section we now discuss
the parametrization (3.12) in a more detail.  Consider the
expectation value of the angular momentum operator $\hat J$ in the normalized
coherent state.  On using (2.8), (3.17), (3.18) and (3.29a)
we find in the boson case
\begin{equation}
\frac{\langle \xi|\hat J|\xi\rangle}{\langle
\xi|\xi\rangle}\equiv \frac{\langle l,\varphi|\hat
J|l,\varphi\rangle}{\langle
l,\varphi|l,\varphi\rangle}=\frac{1}{2\theta_3(\hbox{$
\frac{il}{\pi}$}|\hbox{$\frac{i}{\pi}$})}\frac{d}{dl}\theta_
3(\hbox{$\frac{il}{\pi}$}|\hbox{$\frac{i}{\pi}$}).
\end{equation}
Hence, taking into account the identity
\begin{equation}
\theta_3(\hbox{$\frac{il}{\pi}$}|\hbox{$\frac{i}{\pi}$}) =
\sqrt{\pi}e^{l^2}\theta_3(l|i\pi),
\end{equation}
which follows directly from the general formula
\begin{equation}
\theta_3(\hbox{$\frac{v}{\tau}$}|-\hbox{$\frac{1}{\tau}$}) =
\sqrt{\hbox{$\frac{\tau}{i}$}}\,e^{\frac{i\pi
v^2}{\tau}}\theta_3(v|\tau),
\end{equation}
we arrive at the following relation:
\begin{equation}
\frac{\langle \xi|\hat J|\xi\rangle}{\langle \xi|\xi\rangle}
= l +
\frac{1}{2\theta_3(l|i\pi)}\frac{d}{dl}\theta_3(l|i\pi).
\end{equation}
Now, using the identity [11]
\begin{equation}
\frac{d\theta_3(v)}{dv} =
\pi\theta_3(v)\left(\sum_{n=1}^{\infty}\frac{2iq^{2n-1}e^{2i\pi
v}}{1+q^{2n-1}e^{2i\pi v}} -
\sum_{n=1}^{\infty}\frac{2iq^{2n-1}e^{-2i\pi
v}}{1+q^{2n-1}e^{-2i\pi v}}\right),
\end{equation}
we finally obtain the following equation:
\begin{equation}
\frac{\langle \xi|\hat J|\xi\rangle}{\langle \xi|\xi\rangle}
= l - 2\pi\sin
(2l\pi)\sum_{n=1}^{\infty}\frac{e^{-\pi^2(2n-1)}}{(1+e^{-\pi
^2(2n-1)}e^{2il\pi})(1+e^{-\pi^2(2n-1)}e^{-2il\pi})}\,.
\end{equation}
We note that for $l$ integer or half-integer the above
formula reduces to
\begin{equation}
\frac{\langle \xi|\hat J|\xi\rangle}{\langle \xi|\xi\rangle}
= l,
\end{equation}
that is $l$ is precisely the classical angular momentum.
Otherwise, we observe that the very good approximation of
(3.35) is
\begin{equation}
\frac{\langle \xi|\hat J|\xi\rangle}{\langle \xi|\xi\rangle}
\approx l - 2\pi e^{-\pi^2} \sin (2l\pi).\qquad \hbox{(boson
case)}
\end{equation}
Moreover, it can be easily calculated that the second term
from the right-hand side of (3.37) is still negligible
compared with $l$ (the quantum corrections are really
small).  The maximal error arising in the case $l\to0$ is of
order $0.1$ per cent.  Therefore, in any case
\begin{equation}
\frac{\langle \xi|\hat J|\xi\rangle}{\langle
\xi|\xi\rangle}\approx l.
\end{equation}
We have thus shown that the parameter $l$ in (3.12) can be
identified (in general approximately) with the classical
angular momentum.  In order to demonstrate that the same
holds true in the case with fermions we write down the
identity
\begin{equation}
\theta_2(\hbox{$\frac{il}{\pi}$}|\hbox{$\frac{i}{\pi}$}) =
\sqrt{\pi}e^{l^2}\theta_4(l|i\pi),
\end{equation}
where the Jacobi theta-function $\theta_4$ is defined as
\begin{equation}
\theta_4(v) =
\sum_{n=-\infty}^{\infty}(-1)^nq^{n^2}(e^{i\pi v})^{2n},
\end{equation}
and (3.39) is an immediate consequence of the general
relation
\begin{equation}
\theta_2(\hbox{$\frac{v}{\tau}$}|-\hbox{$\frac{1}{\tau}$}) =
\sqrt{\hbox{$\frac{\tau}{i}$}}\,e^{\frac{i\pi
v^2}{\tau}}\theta_4(v|\tau).
\end{equation}
By (3.39) we can reduce the equation
\begin{equation}
\frac{\langle \xi|\hat J|\xi\rangle}{\langle
\xi|\xi\rangle}\equiv \frac{\langle l,\varphi|\hat
J|l,\varphi\rangle}{\langle
l,\varphi|l,\varphi\rangle}=\frac{1}{2\theta_2(\hbox{$
\frac{il}{\pi}$}|\hbox{$\frac{i}{\pi}$})}\frac{d}{dl}\theta_
2(\hbox{$\frac{il}{\pi}$}|\hbox{$\frac{i}{\pi}$}).
\end{equation}
 following from (2.8), (3.18) with $j$ half-integer, and
(3.29b), to
\begin{equation}
\frac{\langle \xi|\hat J|\xi\rangle}{\langle \xi|\xi\rangle}
= l +
\frac{1}{2\theta_4(l|i\pi)}\frac{d}{dl}\theta_4(l|i\pi).
\end{equation}
Now, taking into account the identity
\begin{equation}
\frac{d\theta_4(v)}{dv} =
\pi\theta_4(v)\left(\sum_{n=1}^{\infty}\frac{2iq^{2n-1}e^{-2i\pi
v}}{1-q^{2n-1}e^{-2i\pi v}} -
\sum_{n=1}^{\infty}\frac{2iq^{2n-1}e^{2i\pi
v}}{1-q^{2n-1}e^{2i\pi v}}\right),
\end{equation}
we finally arrive at the following relation:
\begin{equation}
\frac{\langle \xi|\hat J|\xi\rangle}{\langle \xi|\xi\rangle}
= l + 2\pi\sin
(2l\pi)\sum_{n=1}^{\infty}\frac{e^{-\pi^2(2n-1)}}{(1-e^{-\pi
^2(2n-1)}e^{2il\pi})(1-e^{-\pi^2(2n-1)}e^{-2il\pi})}\,.
\end{equation}
As with the case of bosons we find that for $l$ integer or
half-integer the exact formula is valid such that
\begin{equation}
\frac{\langle \xi|\hat J|\xi\rangle}{\langle \xi|\xi\rangle}
= l.
\end{equation}
Accordingly, for general $l$ we find that the very good
approximation of (3.45) is given by
\begin{equation}
\frac{\langle \xi|\hat J|\xi\rangle}{\langle \xi|\xi\rangle}
\approx l + 2\pi e^{-\pi^2} \sin (2l\pi).\qquad
\hbox{(fermion case)}
\end{equation}
Proceeding as with (3.37) one can easily check that the
approximation (3.47) is as good as
\begin{equation}
\frac{\langle \xi|\hat J|\xi\rangle}{\langle
\xi|\xi\rangle}\approx l.
\end{equation}
It thus appears that analogously as with bosons in the case
of fermions we can also identify the parameter $l$ with the
classical angular momentum.

We now examine the role of the parameter $\varphi$ in
(3.12).  Consider first the case with bosons.  Equations
(3.17), (2.9a), (3.18) and (3.29) taken together yield
\begin{equation}
\frac{\langle \xi|U|\xi\rangle}{\langle \xi|\xi\rangle}
\equiv \frac{\langle l,\varphi|U|l,\varphi\rangle}{\langle
l,\varphi|l,\varphi\rangle} =
e^{-\frac{1}{4}}e^{i\varphi}\,\frac{\theta_2(\hbox{$\frac{il}{
\pi}$}|\hbox{$\frac{i}{\pi}$})}{\theta_3(\hbox{$\frac{il}{
\pi}$}|\hbox{$\frac{i}{\pi}$})}.\qquad \hbox{(boson case)}
\end{equation}
Hence, making use of the relation
\begin{equation}
\theta_2(v) =
e^{i\pi(\frac{\tau}{4}+v)}\theta_3(v+\hbox{$\frac{\tau}{2}$})
\end{equation}
and the identity (3.31) we obtain
\begin{equation}
\frac{\langle \xi|U|\xi\rangle}{\langle \xi|\xi\rangle}
 =
e^{-\frac{1}{4}}e^{i\varphi}\,\frac{\theta_3(l+\hbox{$\frac{1}
{2}$}|i\pi)}{\theta_3(l|i\pi)}.\qquad \hbox{(boson case)}
\end{equation}
Finally, taking into account the definition of the function
$\theta_3$ given by (3.26a) we arrive at the following
formula:
\begin{equation}
\frac{\langle \xi|U|\xi\rangle}{\langle \xi|\xi\rangle}
 =
e^{-\frac{1}{4}}e^{i\varphi}\,\,\frac{\sum\limits_{j=-\infty}^{\infty}e
^{-\pi^2j^2}e^{i\pi(2l+1)j}}{\sum\limits_{j=-\infty}^{\infty}e
^{-\pi^2j^2}e^{i\pi 2lj}}.
\end{equation}
A look at (3.52) is enough to conclude that the very good
approximation of (3.52) is
\begin{equation}
\frac{\langle \xi|U|\xi\rangle}{\langle \xi|\xi\rangle}
 \approx
e^{-\frac{1}{4}}e^{i\varphi}.
\end{equation}
We observe that the approximation (3.53) is valid also in
the case of fermions.  Indeed, using (3.17), (2.9a), (3.18)
and (3.29) we get
\begin{equation}
\frac{\langle \xi|U|\xi\rangle}{\langle \xi|\xi\rangle}
 =
e^{-\frac{1}{4}}e^{i\varphi}\,\frac{\theta_3(\hbox{$\frac{il}{
\pi}$}|\hbox{$\frac{i}{\pi}$})}{\theta_2(\hbox{$\frac{il}{
\pi}$}|\hbox{$\frac{i}{\pi}$})}.\qquad \hbox{(fermion case)}
\end{equation}
Comparing equation (3.54) with (3.49) we find that the
approximation (3.53) holds true.

We now discuss (3.53) in a more detail.  Due to the term
$e^{-\frac{1}{4}}$ in (3.53), it turns out that the average
value of $U$ in the normalized coherent state does not
belong to the unit circle.  In our opinion the reason of
such discrepancy is the inadequate definition of the
expectation value as the diagonal matrix element of $U$.
The fact that average values in some states cannot be
well-defined even in the case of Hermitian operators
(observables) is known in the literature.  We only recall
the expectation value of the usual position operator for the
particle on the real line in its eigenstate.  Following
approach analogous to that described in ref.\ 12 we
introduce the relative average value of $U$ such that
\begin{equation}
\frac{\langle U\rangle_{\xi}}{\langle U\rangle_{\eta}} :=
\frac{\langle \xi|U|\xi\rangle}{\langle \eta|U|\eta\rangle},
\end{equation}
where $|\xi\rangle$ and $|\eta\rangle$ are the normalized
coherent states.  We stress that $\langle U\rangle_{\xi}\ne
\langle \xi|U|\xi\rangle$~.  By virtue of (3.53) (see also
(3.52)) we have
\begin{equation}
\frac{\langle U\rangle_{\xi}}{\langle U\rangle_{\eta}}
\approx e^{i(\varphi-\psi)},
\end{equation}
where $\xi=e^{-l+i\varphi}$ and $\eta=e^{-h+i\psi}$.  In
particular, we find (see (3.21))
\begin{equation}
\frac{\langle U\rangle_{\xi}}{\langle U\rangle_1}
\approx e^{i\varphi}.
\end{equation}
The form of (3.57) suggests that the relative expectation
value $\langle U\rangle_{\xi}/\langle U\rangle_1$ is the
most natural one to describe the average position of a
quantum particle on the circle.

We have thus shown that the parameter $\varphi$ can be
regarded as the classical angle.  Notice that in general we
have the approximate formulas (3.38) and (3.57).  In our
opinion, due to some immanent feature of the quantum
mechanics on the circle, the approximation of the classical
phase space expressed by (3.38) and (3.57) is the best
possible one and in this sense the coherent states marked
with points of such slightly deformed phase space are
closest to the classical ones.

We end this section with discussion of the time evolution of
expectation values of operators $\hat J$, $U$ and $X$ in the
coherent state.  We confine ourselves to the free motion in
the circle.  Evidently, the Hamiltonian is given by (see
(2.3))
\begin{equation}
\hat H = \frac{1}{2}\hat J^2.
\end{equation}
Taking into account (3.58) we get
\begin{equation}
\hat J(t) = \hat J,
\end{equation}
that is $\hat J$ is the integral of motion.  Therefore, by
virtue of (3.38)
\begin{equation}
\frac{\langle \xi|\hat J(t)|\xi\rangle}{\langle
\xi|\xi\rangle}\approx l.
\end{equation}
Thus, as expected, the angular momentum $l$ labelling the
coherent state $|\xi\rangle$ is conserved during the time
evolution.  Furthermore, using the identity which follows
immediately from (2.7)
\begin{equation}
f(\hat J)U = Uf(\hat J + 1),
\end{equation}
where $f$ is an arbitrary analytic function, we find
\begin{equation}
U(t) = e^{it\frac{\hat J^2}{2}}Ue^{-it\frac{\hat
J^2}{2}}=Ue^{it(\hat J + \frac{1}{2})}.
\end{equation}
Now, we have the approximate formula (see appendix)
\begin{equation}
\frac{\langle \xi|e^{s\hat J}|\xi\rangle}{\langle
\xi|\xi\rangle} \approx e^{\frac{1}{4}s^2 + sl},
\end{equation}
and the relation
\begin{equation}
U^\dagger |\xi\rangle = \xi^{-1}e^{-\hat J - \frac{1}{2}}|\xi\rangle.
\end{equation}
The equation (3.64) is implied by (3.4) and
\begin{equation}
X^\dagger |\xi\rangle = \xi^{-1}e^{-2\hat J -1}|\xi\rangle,
\end{equation}
which is a direct consequence of the identity (see (3.4))
\begin{equation}
X^\dagger X = e^{-2\hat J - 1}
\end{equation}
and (3.15).  Eqs.\ (3.62)--(3.64) taken together yield
\begin{equation}
\frac{\langle \xi|U(t)|\xi\rangle}{\langle \xi|\xi\rangle}
 \approx
e^{-\frac{t^2}{4}}e^{-\frac{1}{4}}e^{i(\varphi + tl)}.
\end{equation}
Reasoning as with (3.53) we introduce the relative
expectation value of $U(t)$ of the form
\begin{equation}
\frac{\langle U(t)\rangle_{\xi}}{\langle U(t)\rangle_{\eta}} :=
\frac{\langle \xi|U(t)|\xi\rangle}{\langle \eta|U(t)|\eta\rangle},
\end{equation}
where $|\xi\rangle$ and $|\eta\rangle$ are normalized
coherent states, which in view of (3.67) leads to (see also
(3.57))
\begin{equation}
\frac{\langle U(t)\rangle_{\xi}}{\langle U(t)\rangle_1}
\approx e^{i(\varphi + tl)}.
\end{equation}
Notice that the phase in the exponential function from the
right-hand side of (3.69) is nothing but the solution of the
classical equations of the free motion in the circle such
that
\begin{eqnarray}
\dot \varphi &=& l,\nonumber\\
\dot l &=&0.
\end{eqnarray}

We now discuss the average value of $X(t)$.  Making use of
(3.4) and (3.62) we obtain
\begin{equation}
X(t) = e^{it(\hat J - \frac{1}{2})}X.
\end{equation}
Hence, taking into account (3.1) and (3.63) we get
\begin{equation}
\frac{\langle \xi|X(t)|\xi\rangle}{\langle \xi|\xi\rangle}
 \approx
e^{-\frac{t^2}{4}}e^{-l + i(\varphi + t(l-\frac{1}{2}))}.
\end{equation}
It follows immediately from (3.72) that the corresponding
relative expectation value is
\begin{equation}
\frac{\langle X(t)\rangle_{\xi}}{\langle X(t)\rangle_1}
\approx e^{-l + i(\varphi + t(l-\frac{1}{2}))}.
\end{equation}
On making use of the identification (see (3.12))
\begin{equation}
\xi(t) = e^{-l(t) + i\varphi(t)},
\end{equation}
we find that the argument of the exponential function from
(3.73) refers to the solution of the system
\begin{eqnarray}
\dot \varphi &=& l - \frac{1}{2},\nonumber\\
\dot l &=& 0.
\end{eqnarray}
Evidently, (3.75) reduces to (3.70) in the semiclassical
limit $|l|\gg1$.

We finally point out that the coherent states are not stable
with respect to the free evolution (the standard coherent
states are clearly also unstable in the case with the free
evolution --- recall spreading of wave packets).  Indeed,
using (3.71), (2.8), (3.1) and (3.17) we arrive at the
following formula:
\begin{equation}
X(t)|l,\varphi\rangle = e^{-l +
i(\varphi-\frac{t}{2})}|l,\varphi-t\rangle.
\end{equation}
The natural dynamics for the coherent states which ensures
their stability is that described by the Hamiltonian of the
form
\begin{equation}
\hat H = \omega \hat J.
\end{equation}
In fact, we have
\begin{equation}
U(t) = e^{it\omega\hat J}Ue^{-it\omega\hat J} =
e^{it\omega}U.
\end{equation}
Hence, by virtue of (3.4)
\begin{equation}
X(t) = e^{it\omega}X,
\end{equation}
which leads to
\begin{equation}
X(t)|\xi\rangle = \xi(t)|\xi\rangle,
\end{equation}
where $\xi(t)=e^{it\omega}\xi$.
\subsection{Minimalization of the Heisenberg uncertainty
relations}
In this section we show that the introduced coherent states
minimalize the Heisenberg uncertainty relations.  Let us
define the following Hermitian operators:
\numparts
\begin{eqnarray}
Q &:=& \frac{1}{2}(X + X^\dagger ),\\
P &:=& \frac{1}{2i}(X - X^\dagger ),
\end{eqnarray}
\endnumparts
where $X$ is given by (3.4).  Notice that the operators $Q$
and $P$ play the same role as the usual position and
momentum operators in the case with the standard coherent
states.  Consider the Heisenberg uncertainty relation
\begin{equation}
\Delta_\phi Q\Delta_\phi P \ge \frac{1}{2}\frac{\langle
\phi|[Q,P]|\phi\rangle}{\langle \phi|\phi\rangle}.
\end{equation}
Taking into account (3.81) and the identity
\begin{equation}
XX^\dagger = e^2X^\dagger X,
\end{equation}
where $e=\exp(1)$, which can be derived easily by means of
(3.66), (3.4) and (3.61), we obtain
\begin{equation}
[Q,P] = -\frac{1}{2i}(e^2-1)X^\dagger X.
\end{equation}
By virtue of (3.84) and (3.1) we have
\begin{equation}
\frac{\langle \xi|[Q,P]|\xi\rangle}{\langle \xi|\xi\rangle}
= -\frac{1}{2i}(e^2-1)\xi^*\xi=-\frac{1}{2i}(e^2-1)e^{-2l}.
\end{equation}
Now, equations (3.81), (3.83) and (3.12) taken together yield
\numparts
\begin{eqnarray}
\frac{\langle \xi|Q|\xi\rangle}{\langle \xi|\xi\rangle}
&=& \frac{1}{2}(\xi + \xi^*) = e^{-l}\cos\varphi,\\
\frac{\langle \xi|P|\xi\rangle}{\langle \xi|\xi\rangle}   &=&
\frac{1}{2i}(\xi - \xi^*) = e^{-l}\sin\varphi,\\
\frac{\langle \xi|Q^2|\xi\rangle}{\langle \xi|\xi\rangle}
&=& \frac{1}{4}[(\xi+\xi^*)^2 + (e^2-1)\xi^*\xi] =
\frac{1}{4}e^{-2l}[2\cos2\varphi + e^2 + 1],\\
\frac{\langle \xi|P^2|\xi\rangle}{\langle \xi|\xi\rangle}
&=& -\frac{1}{4}[(\xi-\xi^*)^2 - (e^2-1)\xi^*\xi] =
-\frac{1}{4}e^{-2l}[2\cos2\varphi -(e^2 + 1)].
\end{eqnarray}
\endnumparts
As a consequence of (3.85) and (3.86) we find
\begin{equation}
\Delta_\xi Q\Delta_\xi P = \frac{1}{2}\frac{\langle
\xi|[Q,P]|\xi\rangle}{\langle \xi|\xi\rangle}.
\end{equation}
That is the (normalized) coherent states minimalize the
Heisenberg uncertainty relation (3.82).  We point out that
such minimalization does not take place in the case of the
vectors $|j\rangle$ (see (2.8)).  Indeed, a quick
calculation based on (3.5) and the identity (3.66) shows
that this is not the case.
\section{Bargmann representation}
\subsection{Inner product}
This section is devoted to the Bargmann representation
within the introduced formalism.  We now return to eq.\
(3.18).  Taking into account (3.18) and (2.13) we find that
the resolution of the identity for the coherent states can
be written as
\begin{equation}
\frac{1}{2\pi^{\frac{3}{2}}}\int\limits_{0}^{2\pi}d\varphi
\int\limits_{-\infty}^{\infty}dl\,e^{-l^2}|l,\varphi\rangle
\langle l,\varphi| = I.
\end{equation}
On using the complex variable $\xi$ (see (3.12)) the
completeness condition (4.1) takes the form
\begin{equation}
\frac{1}{4i\pi^{\frac{3}{2}}}\int\limits_{{\bf C}}d\xi
d\xi^*\,\frac{e^{-(\ln|\xi|)^2}}{|\xi|^2}|\xi\rangle\langle
\xi| = I,
\end{equation}
where {\bf C} is the complex plane.  We point out that there
is no singularity in (4.2) for $\xi=0$.  Hence, we arrive at
the functional representation such that
\begin{equation}
\langle \phi|\psi\rangle =
\frac{1}{4i\pi^{\frac{3}{2}}}\int\limits_{{\bf C}}d\xi
d\xi^*\,\frac{e^{-(\ln|\xi|)^2}}{|\xi|^2}\,(\phi(\xi^*))^*\psi
(\xi^*),
\end{equation}
where $\phi(\xi^*)=\langle \xi|\phi\rangle$.  The
representation (4.3) is the counterpart of the Bargmann
representation [13] applied in the theory of standard coherent
states.  The operators act in the representation (4.3) as
follows:
\begin{eqnarray}
\hat J\phi(\xi^*) &=& -\xi^*\frac{d}{d\xi^*}\phi(\xi^*),\\
U\phi(\xi^*) &=& \frac{\phi(e\xi^*)}{\sqrt{e}\,\xi^*},\\
U^\dagger \phi(\xi^*) &=&
\frac{\xi^*}{\sqrt{e}}\phi(e^{-1}\xi^*),\\
T\phi(\xi^*) &=& (\phi(\xi^{-1}))^*,\\
X\phi(\xi^*) &=& \frac{\phi(e^2\xi^*)}{e\xi^*},\\
X^\dagger \phi(\xi^*) &=& \xi^*\phi(\xi^*),
\end{eqnarray}
where $e=\exp(1)$.  In fact the relation (4.4) follows
directly from (3.11) and (3.12).  In order to derive (4.5),
(4.6) and (4.8) we have used the formulas (3.64) and (3.65)
as well as the identities
\begin{eqnarray}
e^{\lambda\xi^*\frac{d}{d\xi^*}} &=&
\sum_{n=0}^{\infty}\frac{(e^\lambda-1)^n}{n!}\xi^{*n}\frac{d
^n}{d\xi^{*n}},\\
\phi(\lambda\xi^*) &=&
\sum_{n=0}^{\infty}\frac{(\lambda-1)^n}{n!}\xi^{*n}\frac{d^n}{d\xi^{
*n}}\phi(\xi^*),
\end{eqnarray}
implying
\begin{equation}
\phi(e^\lambda\xi^*) =
e^{\lambda\xi^*\frac{d}{d\xi^*}}\phi(\xi^*).
\end{equation}
We remark that the relation (4.10) is well-known in the
boson calculus [14].  Finally, the equations (4.7) and (4.9)
are immediate consequences of (3.1) and (3.16),
respectively.

What are the elements $\phi(\xi^*)$ of the Hilbert space
specified by (4.3) ?  In view of (3.10) the basis vectors
$|j\rangle$ are represented by the functions
\begin{equation}
e_j(\xi^*) \equiv \langle \xi|j\rangle =
e^{-\frac{j^2}{2}}\xi^{*-j}.
\end{equation}
Using (4.5) and (4.6) it can be checked easily that
\numparts
\begin{eqnarray}
Ue_j(\xi^*) &=& e_{j+1}(\xi^*),\\
U^\dagger e_j(\xi^*) &=& e_{j-1}(\xi^*).
\end{eqnarray}
\endnumparts
Reasoning as in section 2 we arrive at the representation
generated by means of (4.14) from the vacuum vector
\begin{equation}
e_0(\xi^*) \equiv 1,
\end{equation}
in the case with bosons, and the representation for fermions
with the vacuum vector given by
\begin{equation}
e_{\frac{1}{2}}(\xi^*) = e^{-\frac{1}{4}}\xi^{*-\frac{1}{2}}.
\end{equation}
Clearly, the space of the irreducible representation is the
Hilbert space with the inner product (4.3) of functions
possessing the Laurent series expansion such that
\begin{equation}
\phi(\xi^*) = \sum_{j=-\infty}^{\infty}c_j\xi^{*-j},\qquad
\hbox{(boson case)}
\end{equation}
in the case of bosons, and the space of functions of the
form
\begin{equation}
\phi(\xi^*) =
\sum_{j=-\infty}^{\infty}c_j\xi^{*-j+\frac{1}{2}},\qquad
\hbox{(fermion case)}
\end{equation}
in the case with fermions.

Observe that the expressions from the right-hand side of
(4.5), (4.6) and (4.8) have the structure similar to
$q$-derivatives in the theory of quantum groups.  Such
resemblance is not an accidental one.  Indeed, return to the
equation (3.83).  The formula (3.83) shows that $X$ and
$X^\dagger $ are nothing but the generators of the complex
Manin's plane with involution [15].  On the other hand, we
have the commutation relations
\numparts
\begin{eqnarray}
&&[X,X^\dagger ] = 2\sinh1\,\, e^{-2\hat J},\\
&&[\hat J,X] = X,\qquad [\hat J,X^\dagger ] = -X^\dagger .
\end{eqnarray}
\endnumparts
Hence, introducing the operator $N$ of the form
\begin{equation}
N := -\hat J + \frac{1}{2}\ln(2\sinh1),
\end{equation}
we arrive at the following algebra:
\numparts
\begin{eqnarray}
&&[X,X^\dagger ] = (e^2)^N,\\
&&[N,X] =-X,\qquad [N,X^\dagger ] = X^\dagger .
\end{eqnarray}
\endnumparts
Thus, it turns out that we deal with deformed Heisenberg
algebra.  Notice that the role of the deformation parameter
$q$ is played in both formulas (3.83) and (4.21) by $e^2$.
Therefore, the actual treatment provides us with the concrete
value of $q$.  As pointed out by a referee, on defining
\begin{equation}
a = (1+q)^{-\frac{1}{2}}X,\qquad a^\dagger =
(1+q)^{-\frac{1}{2}}X^\dagger ,
\end{equation}
we can write (4.21) in the form of the usual $q$-boson
algebra.  Indeed, using (4.21) and (3.83) we find
\begin{equation}
aa^\dagger - qa^\dagger a = q^{-N},\qquad [N,a]=-a,\qquad
[N,a^\dagger ]=a^\dagger ,
\end{equation}
where $q=e^{-2}$.  The representations of the algebra (4.23) 
were investigated in [16].  The formulas obtained therein in 
the case of $q<1$ can be regarded as the 
counterparts of the relations (2.8), (3.5) and (3.83) with $X$ 
expressed with the help of (4.22) by $a$, and $\hat J$ given by 
(4.20).  Nevertheless, the physical interpretation of the 
representations of (4.23) in the context of the quantum 
mechanics on the circle was not discussed in [16].  The coherent 
states for the algebra
(4.23) were studied in [17].  Nevertheless, it was assumed
therein that as with standard coherent states the
eigenvalues of the number operator $N$ are nonnegative
integers.  On the contrary, the eigenvalues of the operator
$N$ given by (4.20) can be also negative and non-integer.  We
conclude that the coherent states discussed in [17] are
different from those studied in this work.
\subsection{Reproducing kernel}
Recall that the existence of the reproducing kernel is one
of the most characteristic properties of the standard
Bargmann representation and its numerous generalizations.
From (4.2) and (3.24) it follows immediately that the
reproducing property can be written in the form
\begin{equation}
\phi(\eta^*) =
\frac{1}{4i\pi^{\frac{3}{2}}}\int\limits_{{\bf C}}d\xi
d\xi^*\,\frac{e^{-(\ln|\xi|)^2}}{|\xi|^2}\,{\cal
K}_{0,\frac{1}{2}}(\eta^*,\xi)\phi(\xi^*),
\end{equation}
where $\phi(\eta^*)=\langle \eta|\phi\rangle$ and the indices
0 and $\frac{1}{2}$
correspond to the case with bosons and fermions,
respectively.  The reproducing kernels ${\cal
K}_{0,\frac{1}{2}}(\eta^*,\xi)$ are given by
\numparts
\begin{eqnarray}
{\cal K}_0(\eta^*,\xi) &=&
\theta_3(\hbox{$\frac{i}{2\pi}$}\ln\eta^*\xi|\hbox{$\frac{i}
{\pi}$}),\qquad \hbox{(boson case)}\\
{\cal K}_{\frac{1}{2}}(\eta^*,\xi) &=&
\theta_2(\hbox{$\frac{i}{2\pi}$}\ln\eta^*\xi|\hbox{$\frac{i}
{\pi}$}).\qquad \hbox{(fermion case)}
\end{eqnarray}
\endnumparts
In particular, using (4.2) and (3.24) we get the following
identities satisfied by the theta-functions $\theta_3$ and
$\theta_2$:
\numparts
\begin{eqnarray}
\theta_3(\hbox{$\frac{i}{2\pi}$}\ln\xi^*\eta|\hbox{$\frac{i}
{\pi}$}) &=& \frac{1}{4i\pi^{\frac{3}{2}}}\int\limits_{{\bf
C}}d\zeta d\zeta^*\,\frac{e^{-(\ln|\zeta|)^2}}{|\zeta|^2}\,
\theta_3(\hbox{$\frac{i}{2\pi}$}\ln\xi^*\zeta|\hbox{$\frac{i}
{\pi}$})
\theta_3(\hbox{$\frac{i}{2\pi}$}\ln\zeta^*\eta|\hbox{$\frac{i}
{\pi}$}),\\
\theta_2(\hbox{$\frac{i}{2\pi}$}\ln\xi^*\eta|\hbox{$\frac{i}
{\pi}$}) &=& \frac{1}{4i\pi^{\frac{3}{2}}}\int\limits_{{\bf
C}}d\zeta d\zeta^*\,\frac{e^{-(\ln|\zeta|)^2}}{|\zeta|^2}\,
\theta_2(\hbox{$\frac{i}{2\pi}$}\ln\xi^*\zeta|\hbox{$\frac{i}
{\pi}$})
\theta_2(\hbox{$\frac{i}{2\pi}$}\ln\zeta^*\eta|\hbox{$\frac{i}
{\pi}$}).
\end{eqnarray}
\endnumparts
\newpage
\noindent Taking into account (4.1) and (3.12) we arrive at the
equivalent form of (4.26) such that
$$\displaylines{
\hspace{.5em}\theta_3(\hbox{$\frac{1}{2\pi}$}(\varphi-\psi)-\hbox{$\frac{
l+h}{2}\frac{i}{\pi}$}|\hbox{$\frac{i}{\pi}$})\hfill\cr
\hspace{1.5em}=\frac{1}{2\pi^{\frac{3}{2}}}\int\limits_{0}^{2\pi}d\alpha
\int\limits_{-\infty}^{\infty}dp\,e^{-p^2}
\theta_3(\hbox{$\frac{1}{2\pi}$}(\varphi-\alpha)-\hbox{$\frac{
l+p}{2}\frac{i}{\pi}$}|\hbox{$\frac{i}{\pi}$})
\theta_3(\hbox{$\frac{1}{2\pi}$}(\alpha-\psi)-\hbox{$\frac{
h+p}{2}\frac{i}{\pi}$}|\hbox{$\frac{i}{\pi}$}),\hfill\llap{(4.27a)}\cr
\hspace{.5em}\theta_2(\hbox{$\frac{1}{2\pi}$}(\varphi-\psi)-\hbox{$\frac{
l+h}{2}\frac{i}{\pi}$}|\hbox{$\frac{i}{\pi}$})\hfill\cr
\hspace{1.5em}=\frac{1}{2\pi^{\frac{3}{2}}}\int\limits_{0}^{2\pi}d\alpha
\int\limits_{-\infty}^{\infty}dp\,e^{-p^2}
\theta_2(\hbox{$\frac{1}{2\pi}$}(\varphi-\alpha)-\hbox{$\frac{
l+p}{2}\frac{i}{\pi}$}|\hbox{$\frac{i}{\pi}$})
\theta_2(\hbox{$\frac{1}{2\pi}$}(\alpha-\psi)-\hbox{$\frac{
h+p}{2}\frac{i}{\pi}$}|\hbox{$\frac{i}{\pi}$}).\hfill\llap{(4.27b)}\cr}$$
\setcounter{equation}{27}%
The authors have not found integral identities for the
theta-functions like (4.26) and (4.27) in the literature.
\subsection{The action of the operators}
Finally, we study the action of operators in the Bargmann
representation.  Let $A$ be a linear operator.  Making use
of (4.2) we get
\begin{equation}
(A\phi)(\eta^*) =
\frac{1}{4i\pi^{\frac{3}{2}}}\int\limits_{{\bf C}}d\xi
d\xi^*\,\frac{e^{-(\ln|\xi|)^2}}{|\xi|^2}\,{\cal
A}(\eta^*,\xi)\phi(\xi^*).
\end{equation}
Equation (4.28) implies that an arbitrary linear operator $A$
is represented in the Bargmann representation by the integral
operator.  Taking into account (2.14) and (3.10) we find
that the kernel ${\cal A}(\eta^*,\xi)$ of the integral
operator (4.28) can be written as
\begin{equation}
{\cal A}(\eta^*,\xi) =
\sum_{j,k=-\infty}^{\infty}A_{jk}\,\eta^{*-j}\xi^{-k}e^{-\frac
{1}{2}(j^2+k^2)},
\end{equation}
where $A_{jk}=\langle j|A|k\rangle$.  On the other hand, we
have
\begin{equation}
{\cal A}(\eta^*,\xi) = \langle \eta|A|\xi\rangle .
\end{equation}
Therefore, the kernel ${\cal A}$ is linked to the covariant
symbol
\begin{equation}
A(\xi^*,\xi) = \frac{\langle
\xi|A|\xi\rangle}{\langle \xi|\xi\rangle}
\end{equation}
of the operator $A$ by
\begin{eqnarray}
{\cal A}(\xi^*,\xi) &=& \langle \xi|\xi\rangle
A(\xi^*,\xi)\nonumber\\
&=& \left\{
\begin{array}{ll}
\theta_3(\hbox{$\frac{i}{\pi}$}\ln|\xi||\hbox{$\frac{i}
{\pi}$})A(\xi^*,\xi),
 & \hbox{(boson case)}\\
\theta_2(\hbox{$\frac{i}{\pi}$}\ln|\xi||\hbox{$\frac{i}
{\pi}$})A(\xi^*,\xi).
 & \hbox{(fermion case)}
\end{array}
\right.
\end{eqnarray}

Consider now the linear operators $K_0$ and
$K_{\frac{1}{2}}$ defined as
\begin{equation}
(K_{0,\frac{1}{2}}\phi)(\eta^*) :=
\frac{1}{4i\pi^{\frac{3}{2}}}\int\limits_{{\bf C}}d\xi
d\xi^*\,\frac{e^{-(\ln|\xi|)^2}}{|\xi|^2}\,{\cal
K}_{0,\frac{1}{2}}(\eta^*,\xi)\phi(\xi^*),
\end{equation}
where ${\cal K}_0$ and ${\cal K}_{\frac{1}{2}}$ are the
reproducing kernels given by (4.25).  Evidently, the
operators $K_0$ and $K_{\frac{1}{2}}$ are the projectors on
the space of the irreducible representation of the algebra
(2.7), (2.10) and (2.11) corresponding to the case with
$j_0=0$ (bosons) and $j_0=\frac{1}{2}$ (fermions),
respectively.  Therefore, we can formalize the quantization
procedure by demanding that
\begin{equation}
(K_0\phi)(\xi^*) = \phi(\xi^*),\qquad \hbox{(boson case)}
\end{equation}
in the boson case, and
\begin{equation}
(K_{\frac{1}{2}}\phi)(\xi^*) = \phi(\xi^*),\qquad
\hbox{(fermion case)}
\end{equation}
in the case with fermions.  Notice that the resolution of
the identity (4.2) is rather formal.  Actually, we have

\begin{equation}
\int\limits_{[0,1]}\!\!\raise10pt\hbox{$\scriptstyle\oplus$}
dj_0\,K_{j_0} = I,
\end{equation}
where $\int^{\oplus}$ designates the direct integral.
\section{Example: free motion in the circle}
We now illustrate the introduced formalism by
an example of the Schr\"o\-dinger equation for the free particle
on the circle.  We restrict for brevity
to the case with bosons.  The corresponding Hamiltonian is
\begin{equation}
\hat H = \frac{1}{2} \hat J^2.
\end{equation}
By virtue of (2.8) the normalized solution of the
Schr\"odinger equation
\begin{equation}
\hat H |E\rangle = E|E\rangle,
\end{equation}
is given by
\begin{equation}
|E\rangle = |j\rangle,\qquad E = \frac{1}{2}j^2,
\end{equation}
where $j$ is integer, that is $\hat H$ has purely discrete
simple spectrum.  It must be borne in mind that due to the
twofold degeneracy the vector $|-j\rangle$ is also the
normalized solution to (5.2).  Clearly, both the solutions
are related by (2.12).  The fact that the operator of the
time inversion $T$ maps the solution $|j\rangle$ into
another solution $|-j\rangle$ of the Schr\"odinger equation
(5.2) is a direct consequence of the $T$-invariance of the
Hamiltonian (5.1) (see (2.10)).  Notice that (3.10) implies
the following form of the solution (5.3) in the Bargmann
representation:
\begin{equation}
\phi_j(\xi^*) \equiv \langle \xi|E\rangle =
e^{-\frac{j^2}{2}}\xi^{*-j}.
\end{equation}

We now discuss the distribution of energies in the coherent
state.  Taking into account (3.10), (3.28a) and (3.12) we
find
\begin{equation}
\frac{|\langle j|\xi\rangle|^2}{\langle \xi|\xi\rangle} =
\frac{|\xi|^{-2j}e^{-j^2}}{\theta_3(\hbox{$\frac{i}{\pi}$}
\ln|\xi||\hbox{$\frac{i}{\pi}$})} = \frac{e^{2lj}e^{-j^2}}{
\theta_3(\hbox{$\frac{il}{\pi}$}|\hbox{$\frac{i}{\pi}$})}.
\end{equation}
Referring back to eq.\ (3.31) and making use of the
definition of the theta-function $\theta_3$ (3.26a) we
arrive at the approximate relation
\begin{equation}
\theta_3(\hbox{$\frac{il}{\pi}$}|\hbox{$\frac{i}{\pi}$}) =
\sqrt{\pi}\,e^{l^2}\left(1 +
2\sum_{n=1}^{\infty}e^{-\pi^2n^2}\cos(2\pi ln)\right)
\approx \sqrt{\pi}\,e^{l^2}.
\end{equation}
Using this we can write (5.5) in the following form:
\begin{equation}
\frac{|\langle j|\xi\rangle|^2}{\langle \xi|\xi\rangle}
\approx \frac{1}{\sqrt{\pi}}e^{-(j-l)^2}.
\end{equation}
We have thus shown that the probability of finding the
system in the state $|j\rangle$ when the system is in the
normalized coherent state $|\xi\rangle/\sqrt{\langle
\xi|\xi\rangle}$ has a ``discrete'' Gaussian distribution.
Recall that the counterpart of (5.7) in the case of the
harmonic oscillator is the asymmetric Poisson distribution.
\section{Conclusion}
We have introduced in this work the coherent states for the
quantum particle on the circle.  An advantage of the actual
approach is that it provides a concrete representation of
the algebra (2.7) which seems to be the most natural one for
description of the circular motion of a quantum particle.
Rather unexpected result is the appearance of the deformed
Heisenberg algebra (4.21) in the introduced formalism.
There are some indications that this observation would
provide some insight into the physical interpretation of
quantum deformations.  Namely, it seems that the deformed
algebra (4.21) would be connected with the lattice-like
structure on the classical phase space (the cylinder)
implied by the exact relations (3.36) and (3.46) for $l$
integer or half-integer.  This idea is strongly supported by
the recent observations of Celeghini \etal [18] who demonstrated
that in the Fock-Bargmann representation of the quantum
Weyl-Heisenberg algebra in the space of analytic (entire)
functions, the coherent and squeezed states and the quantum
algebra itself are naturally related to discretized
(periodic) quantum systems with the lattice spacing
associated with the deformation parameter.  Last but not least, 
we point out that results of this paper would be of importance in 
the theory of special functions.  Indeed, the integral
identities for the theta-functions (4.26) and (4.27) are
most probably new.
\ack
This work was supported by KBN grant 2P30221706p01.  We
would like to thank referees for helpful comments.
\appendix
\section{}
We append the proof of the following approximate relation:
\begin{equation}
\frac{\langle \xi|e^{s\hat J}|\xi\rangle}{\langle
\xi|\xi\rangle} \approx e^{\frac{1}{4}s^2 + sl}.
\end{equation}
{\em Proof}.\quad By (2.8), (3.17), (3.18) and (3.29a) (see
also (3.30)) we have
\begin{equation}
\frac{\langle \xi|{\hat J}^k|\xi\rangle}{\langle
\xi|\xi\rangle} =
\frac{1}{2^k}\frac{1}{\theta_3}\frac{d^k\theta_3}{dl^k}.
\end{equation}
Furthermore, as we have already shown (see (3.30)) and
(3.38)) the following approximate formula holds:
\begin{equation}
\frac{1}{\theta_3}\frac{d\theta_3}{dl} \approx
2l.
\end{equation}
Now, consider the identity
\begin{equation}
\frac{1}{\theta_3}\frac{d^{k+1}\theta_3}{dl^{k+1}} =
\frac{1}{\theta_3^2}\frac{d\theta_3}{dl}\frac{d^k\theta_3}{d
l^k} +
\frac{d}{dl}\left(\frac{1}{\theta_3}\frac{d^k\theta_3}{dl^k}
\right),
\end{equation}
where $k=0,\,1,\,2,\,\ldots$.  In view of (A.3) we can write
(A.4) in the form of the recursive relation
\begin{equation}
f_{k+1} = 2lf_k + \frac{df_k}{dl},\qquad
f_0\equiv1,
\end{equation}
where
\begin{equation}
f_k = \frac{1}{\theta_3}\frac{d^k\theta_3}{dl^k},\qquad
k=0,\,1,\,2,\,\ldots\,.
\end{equation}
We note that due to (A.3) the equation (A.5) is valid only
approximately.  Evidently,
\begin{equation}
\frac{\langle \xi|e^{s\hat J}|\xi\rangle}{\langle
\xi|\xi\rangle} =
\sum_{k=0}^{\infty}\frac{1}{2^k}\frac{f_k}{k!}s^k.
\end{equation}
Let us introduce the function $u(s,l)$ such that
\begin{equation}
u(s,l) = \frac{\langle \xi|e^{s\hat J}|\xi\rangle}{\langle
\xi|\xi\rangle}
\end{equation}
From (A.5) and (A.7) it follows that
\begin{equation}
\frac{\partial u(s,l)}{\partial s} = lu(s,l) +
\frac{1}{2}\frac{\partial u}{\partial l},\qquad
u(0,l)\equiv1.
\end{equation}
The solution to (A.9) is
\begin{equation}
u(s,l) = e^{\frac{1}{4}s^2 +
sl}.
\end{equation}
This observation completes the proof.  We note that the
proof refers to the case of bosons.  Nevertheless, it
can be easily checked (see (3.42) and (3.48)) that the
identity (A.1) holds true for fermions as well.\\[\baselineskip]
{\em Example}: For an easy illustration of (A.1)
consider the identity (3.66)
$$\displaylines{
\hfill X^\dagger X = e^{-2\hat J - 1}.\hfill\cr}$$
On the one hand, using (3.1) and (3.12) we find
\begin{equation}
\frac{\langle \xi|X^\dagger X|\xi\rangle}{\langle
\xi|\xi\rangle} = e^{-2l}.
\end{equation}
On the other hand, (A.1) implies
\begin{equation}
\frac{\langle \xi|e^{-2\hat J -1}|\xi\rangle}{\langle
\xi|\xi\rangle} \approx e^{-2l}.
\end{equation}
That is in the discussed case the formula (A.1) provides us
with the exact result.


\newpage
\section*{References}
\begin{thebibliography}{VV}
\bibitem{}
DeWitt-Morette C, private communication
\bibitem{}
Ali M K 1995 The double pendulum (unpublished)
\bibitem{}
Emch G G 1972 {\em Algebraic Methods in Statistical
Mechanics and Quantum Field Theory} (New York: Wiley)
\bibitem{}
De Bi\`evre S and Gonz\'alez J A 1993
Semiclassical behaviour of coherent states on the circle,
in: Ali S T, Ladanov I M and Odzijewicz A, editors, {\em
Quantization and Coherent States Methods in Mathematical
Physics}, Proceedings of 11th Workshop on Geometrical
Methods in Mathematical Physics, Bialystok 1992 (Singapore: World
Scientific)
\bibitem{}
Louisell W H 1963 {\em Phys. Lett.} {\bf 7} 60
\\ Carruthers P and Nieto M M 1968 {\em Rev. Mod. Phys.}
{\bf 40} 411\\
Lynch R 1995 {\em Phys. Rep.} {\bf 256} 367
\bibitem{}
Isham C J 1984 Topological and global
aspects of quantum theory, in: DeWitt B S and Stora R,
editors, {\em Relativity, Groups and Topology}\/ II, pp.
1059--1290, Les Houches, Session XL. 1983, Elsevier
\bibitem{}
Pradhan T and Khare A V 1973 {\em Am. J. Phys.} {\bf 41} 59
\bibitem{}
Ohnuki Y and Kitakado S 1993 {\em J. Math. Phys.} {\bf 34} 2827
\bibitem{}
Perelomov A M 1986 {\em Generalized Coherent States and
Their Applications} (Berlin: Springer)
\bibitem{}
Barut A O and Girardello L 1971 {\em Commun. Math. Phys.} {\bf
21} 41
\bibitem{}
Whittaker E T and Watson G N 1963 {\em A Course of
Modern Analysis} (Cambridge: Cambridge University Press)
\bibitem{}
Bogoliubov N N, Logunov A A and Todorov I T 1975
{\em
Introduction to Axiomatic Quantum Field Theory}
(New York: Benjamin)
\bibitem{}
Bargmann V 1961 {\em Commun. Pure Appl. Math.} {\bf 14} 187
\bibitem{}
Louisell W H 1964 {\em Radiation
and Noise in Quantum Electronics} (New York: McGraw-Hill)
\bibitem{}
Woronowicz S. L. 1991 {\em Commun. Math.
Phys.} {\bf 136} 399\\Brzezinski T, Dabrowski H
and Rembielinski J 1992 {\em J. Math. Phys.} {\bf 33} 19
\bibitem{}
Rideau G 1992 {\em Lett. Math. Phys.} {\bf 24} 147
\bibitem{}
Solomon A I 1994 {\em Phys. Lett. A} {\bf 196} 29\\
Wang J, Chuankui W and Jinyu H 1995 {\em Phys. Lett. A} {\bf 199} 137
\bibitem{}
Celeghini E, De Martino S, De Siena S, Vitiello G and Rasetti M
1993 {\em Mod. Phys. Lett. B} {\bf 20} 1321\\Celeghini E, De Martino S,
De Siena S, Rasetti M and Vitiello G 1995 {\em Ann. of Phys. (N.Y.)}
{\bf 241} 50
\end{thebibliography}
\end{document}